\documentstyle[preprint,aps]{revtex}
\tighten

\begin{document}
\draft
\preprint{\vbox{Submitted to Phys.\ Rev.\ C \hfill FSU-SCRI-94-62}}
\title{Relativistic treatment of spin-transfer observables \\
       in quasielastic $(\vec{\bf p},\vec{\bf n})$ scattering}
\author{C.J. Horowitz}
\address{Nuclear Theory Center and Department of Physics, \\
         Indiana University, Bloomington, IN 47505}
\author{J. Piekarewicz}
\address{Supercomputer Computations Research Institute, \\
         Florida State University, Tallahassee, FL 32306}
\date{\today}
\maketitle
 
\begin{abstract}

We calculate all spin-transfer observables for the quasielastic
$(\vec{p},\vec{n})$ reaction in a relativistic plane-wave 
impulse approximation. The nuclear-structure information is 
contained in a large set of nuclear-response functions that are
computed in nuclear matter using a relativistic random-phase 
approximation to the Walecka model. A reduced value of the 
nucleon mass in the medium induces important dynamical changes 
in the residual isovector interaction relative to its nonrelativistic 
counterpart. As a result, good agreement is found for all spin 
observables --- including the spin-longitudinal to spin-transverse 
ratio --- when compared to the original ($q=1.72$~fm$^{-1}$) NTOF 
experiment. In contrast, the spin-longitudinal to spin-transverse 
ratio is underpredicted at $q=1.2$~fm$^{-1}$ and overpredicted at 
$q=2.5$~fm$^{-1}$. We comment on the role of distortions as a 
possible solution to this discrepancy.
\end{abstract}

\narrowtext

\section{Introduction}
\label{secintro}

The prominence of the pion as a mediator of the nucleon-nucleon (NN) 
force is indisputable. Yet, its propagation through the nuclear medium 
is far from understood and remains the source of considerable debate.
The propagation of a pion through the nuclear medium is modified by 
its coupling to nuclear (e.g., particle-hole) excitations. This 
information is contained in the meson self-energy whose imaginary part 
is a physical observable characterizing the linear response of the system 
to an external probe. Hence, information about pion propagation 
through the nuclear medium may be obtained, in principle, from a 
measurement of the nuclear isovector response. Several different 
experiments have been designed with the aim of extracting the nuclear 
isovector response\cite{carey84,rees86}, yet the development of 
sophisticated experimental facilities and techniques have made the 
$(\vec{p},\vec{n})$ reaction the paradigm~\cite{mccle92,chen93}. 
Indeed, the recent availability of a complete set of 
$(\vec{p},\vec{n})$ spin-transfer observables has made possible
the extraction of the spin-longitudinal response ($R_{L}$). This 
response is important because it is sensitive to the propagation
of the pion through the nuclear medium. Moreover, it is not accessible 
with electromagnetic probes, yet it is as fundamental as the longitudinal 
and transverse responses measured in electron scattering.

Inextricably linked to the pion in the study of the isovector 
response is the $\rho$-meson. Although the heavier $\rho-$meson 
contributes to the more uncertain short-distance dynamics, its role 
as a mediator of the NN interaction is fairly well established. 
Indeed, meson-exchange models of the NN interaction, as well as 
the large isovector anomalous moment of the nucleon, suggest a 
strong tensor coupling of the nucleon to the $\rho$. This 
tensor-coupling is, in particular, responsible for the
regularization of the large pionic contribution to the NN tensor 
force. Moreover, rho exchange dominates the spin-transverse 
component of the isovector interaction and is, thus, ultimately 
responsible for the collective behavior of the spin-transverse 
response ($R_{T}$). This is in contrast to $R_{L}$ which is 
dominated by pion exchange.

The $\pi-\rho$ mass difference makes the study of the momentum-transfer 
dependence of $R_{L}$ and $R_{T}$ particularly interesting. In the 
conventional ($\pi + \rho + g^{'}$) model of the residual interaction
the longitudinal component of the interaction becomes attractive at a 
momentum transfer of $q\simeq 1$ fm$^{-1}$ due to the small pion mass. 
In contrast, the larger mass of the $\rho$ meson causes the transverse 
component to remain repulsive well beyond $q\simeq 2.5$ fm$^{-1}$. 
This behavior lead Alberico, Ericson, and Molinari to predict a
softening and enhancement of the spin-longitudinal response and a 
quenching and hardening of the spin-transverse response~\cite{alber82}. 
Thus, they suggested that a measurement of $R_{L}$ and $R_{T}$ should 
reveal a large enhancement in the ratio ($R_{L}/R_{T}$) at low-energy 
loss relative to its free Fermi-gas value. The lack of an experimental 
enhancement in $R_{L}/R_{T}$ constitutes one of the most serious challenges 
facing nuclear physics today~\cite{carey84,rees86,mccle92,chen93}.

	The measurement of $(\vec{p},\vec{n})$ spin observables in the 
quasielastic region might constitute the cleanest and most unambiguous 
way of probing the nuclear spin-isospin response. First, the reactive 
content of the reaction is dominated by quasifree nucleon knockout. 
Indeed, for momentum transfers in excess of 1 fm$^{-1}$ a clear 
quasielastic peak is seen at an excitation energy close to the 
corresponding one for free NN scattering. Moreover, free NN spin 
observables provide a baseline, against which possible medium 
effects may be inferred. Deviations of spin observables from 
their free NN values are likely to arise from a modification 
of the NN interaction inside the nuclear medium or from a 
change in the collective response of the target. Indeed, relativistic 
calculations predict medium modifications to the free NN interaction 
steming from an enhanced lower component of Dirac spinors in the medium. 
This modified NN interaction is responsible for one of the clearest 
relativistic signature found to date --- the quenching of the analyzing 
power in $(\vec{p},\vec{p}')$ quasielastic scattering~\cite{hormur88}. 
Second, nonrelativistic calculations of $(\vec{p},\vec{n})$ observables 
at $q=1.72$~fm$^{-1}$ have shown that while distortions provide an 
overall reduction of the cross section, they do so without substantially 
modifying the distribution of strength~\cite{ichim89}. Finally, the 
$(\vec{p},\vec{n})$ reaction acts as an isospin filter by isolating 
the isovector component of the response. Thus, a combined measurement 
of $(\vec{p},\vec{n})$ and $(\vec{p},\vec{p}')$ spin observables should
enable one to determine the spin-isospin content of the nuclear 
response.

Recently we have reported the first relativistic calculations of 
$R_{L}/R_{T}$ using a relativistic random-phase approximation (RPA) 
to the Walecka model~\cite{horpie93}. In the Walecka model nucleons 
interact via the exchange of isoscalar $\sigma$ (scalar) and $\omega$
(vector) mesons~\cite{waleck74,serwal86}. In a mean-field approximation 
the scalar and vector meson fields are replaced by their classical 
expectation value. The mean-field approximation is characterized by 
the appearance of strong scalar and vector mean fields which induce 
large shifts in the mass and energy of a particle in the medium. 
Isoscalar effects from a reduced nucleon mass lead to important changes 
in the isovector response. In particular, a reduced nucleon mass is
responsible for a shift in the position and an increase in the width 
of the quasielastic peak that are sufficient to explain the ``quenching'' 
and ``hardening'' of the transverse response. In addition, a smaller 
nucleon mass leads to a significant reduction in the effective $NN\pi$ 
coupling in the nuclear medium~\cite{dawpie91}. This effect reduces the 
enhancement of the spin-longitudinal response relative to nonrelativistic 
predictions and leads to no enhancement in $R_{L}/R_{T}$, in agreement 
with experiment~\cite{mccle92,chen93}. In this work we present a detailed 
account of our assumptions and calculational scheme. Moreover, we extend 
previous results, limited to $R_{L}/R_{T}$, to include all spin-transfer 
observables.

We have organized our paper as follows: In Sec.~\ref{secformal} we 
present the relevant formalism for the calculation of the 
isovector response. This section includes a discussion
of the different assumptions and approximations leading to the 
spin-dependent cross section. The relativistic parameterization 
of the underlying NN t-matrix, the nuclear isovector response, 
and the residual particle-hole interaction are also presented
in this section. In Sec.~\ref{secres} results are presented 
for a variety of approximations ranging from a free Fermi gas 
to a relativistic RPA treatment of the nuclear response. Finally, 
Sec.~\ref{secconcl} contains a summary of our findings and 
conclusions.

\section{Formalism}
\label{secformal}

\subsection{Relativistic plane-wave impulse approximation}
\label{rpwia}

In the impulse approximation the interaction between two nucleons
in the medium is assumed to be identical to their interaction in 
free space. Thus, in the impulse approximation the in-medium NN 
interaction is fully determined from two-nucleon data. A convenient 
representation of the NN amplitude is given in terms of five 
Lorentz-invariant amplitudes. A standard parameterization of
the NN amplitude includes scalar, vector, tensor, pseudoscalar, 
and axial-vector amplitudes~\cite{mrw83}
\begin{eqnarray}
   {\cal F} &=&  {\cal F}_{S} 
             +   {\cal F}_{V} \gamma^{\mu}_{(1)}
                 \gamma^{\phantom{\mu}}_{\mu (2)} 
             +   {\cal F}_{T} \sigma^{\mu\nu}_{(1)}
                 \sigma^{\phantom{\mu\nu}}_{\mu\nu (2)}    \nonumber  \\ 
            &+&  {\cal F}_{P} \gamma^{5}_{(1)}
                              \gamma^{5}_{(2)} 
             +   {\cal F}_{A} \gamma^{\mu}_{(1)}\gamma^{5}_{(1)} 
                 \gamma^{\phantom{\mu}}_{\mu (2)}\gamma^{5}_{(2)} \;,
 \label{relinv}
\end{eqnarray}
where the subscripts $(1)$ and $(2)$ refer to the incident and
struck nucleons, respectively and we will adopt the conventions
of Bjorken and Drell for the gamma matrices~\cite{bjodre65}.
Note that the amplitudes depend on two Lorentz-invariant quantities:
the square of the total four momentum ($s$) and the square of the 
four-momentum transfer ($t$). Matrix elements of ${\cal F}$ taken 
between free Dirac spinors can be regarded as known since they are 
directly related to the free NN phase shifts~\cite{mrw83}.

In the following presentation of the $(\vec{p},\vec{n})$ cross
section we shall use only the scalar invariant, for simplicity. 
Our final results, presented at the end of this section, will 
include all additional Lorentz structures. In a relativistic 
plane-wave impulse approximation (RPWIA) the cross section for 
the charge-exchange process is given by 
\begin{equation}
  d\sigma_{{\bf s}'{\bf s}} = {2\pi \over |{\bf v}|} 
                  \int {d{\bf p'} \over (2\pi)^{3}} |t_{S}|^{2} 
                  L({\bf p}s;{\bf p'}s') S({\bf q},\omega) \;,
 \label{sigmas1}
\end{equation}
where $|{\bf v}|$ is the incident flux in the rest frame of the
nucleus, ${\bf p}({\bf p'})$ and $s(s')$ are the momentum and spin 
projection of the initial(final) nucleon, and ${\bf q}$ and $\omega$
are the momentum and energy transfer to the nucleus, respectively.
All dynamical information about the process is contained 
in the three quantities $t_{S}$, $L$, and $S$. $t_{S}$ is the scalar 
component of the NN t-matrix driving the reaction. The plane-wave
tensor $L({\bf p}s;{\bf p'}s')$ contains information about the 
polarization of the incoming and outgoing nucleon. Finally, 
$S({\bf q},\omega)$ characterizes the nuclear response. We address 
each contribution individually below. 

The scalar component of the NN t-matrix in the center of momentum 
frame is simply related to the corresponding Lorentz invariant 
amplitude~\cite{hormur88}
\begin{equation}
  t_{S} \equiv -8\pi i \;
  {p_{\rm cm} E_{\rm cm} \over M^{2}} {\cal F}_{S} \;.
  \label{tNN}
\end{equation}
In particular, in a one-boson exchange description of the 
scattering process, and to lowest order in the coupling, 
it reduces to:
\begin{equation}
  t_{S} \rightarrow v_{S} =
  {g_{S}^{2} \over q_{\mu}^{2} - m_{S}^{2}} \;,
 \label{vscalar}
\end{equation}
where $q_{\mu}$ is the (spacelike) four-momentum transfer
to the nucleus, and $m_{S}$ and $g_{S}$ are the scalar mass 
and coupling constant, respectively.

The polarization information is contained in the projectile ``tensor'' 
\begin{equation}
  L({\bf p},{\bf s};{\bf p'},{\bf s}') = 
  \Big| \;
    \bar{{\cal U}}({\bf p'},s') \; {\bf 1} \; {\cal U}({\bf p},s) 
  \Big|^{2} \;.
 \label{ltensor1} 
\end{equation}
In a relativistic plane-wave approximation the tensor can be written 
exclusively in terms of free Dirac spinors. In nuclear matter, however, 
the effective mass of a nucleon is reduced relative to its free space 
value by the presence of a scalar mean field. Thus, for a nucleon 
propagating through nuclear matter the in-medium Dirac spinor 
becomes~\cite{serwal86}
\begin{equation}
 {\cal U}({\bf p},s) =
 \sqrt{{ E^{\star}_{\bf p} + M_{p}^{*} \over 2E^{\star}_{\bf p} }}
 \left(
  \begin{array}{c}
            1      \\
      \displaystyle{
        {\sigma\cdot{\bf p} \over 
         E^{\star}_{{\bf p}} + M_{p}^{\star} }}
  \end{array}  
 \right) \chi_{s} \;,
 \label{freesp}
\end{equation}
where $M_{p}^{\star}$ is the effective mass of the projectile
and the on-shell energy is given by
\begin{equation}
 E^{\star}_{\bf p}=\sqrt{{\bf p}^{2}+M_{p}^{\star 2}} \;.
\end{equation}
Note that the normalization implied by Eq.~(\ref{freesp}) differs
from the one in Ref.~\cite{bjodre65} in that
\begin{equation}
  {\cal U}^{\dagger}({\bf p},s)     \;
  {\cal U}({\bf p},s')=\delta_{ss'} \;.
 \label{normal}
\end{equation}
The plane-wave approximation enables one to write, and ultimately
to evaluate, the projectile tensor using Feynman's trace techniques
\begin{equation}
  L({\bf p},{\bf s};{\bf p'},{\bf s}') = 
  {\rm Tr}
  \left[ 
    \left({ {\rlap/{p}} + M_{p}^{\star} \over 2E^{\star}_{\bf p}  }   \right)
    \left({ 1 + \gamma^{5}{\rlap/{s}} \over 2 }   \right)
    \left({ {\rlap/{p'}} + M_{p}^{\star} \over 2E^{\star}_{\bf p'}  } \right)
    \left({ 1 + \gamma^{5}{\rlap/{s'}} \over 2  } \right)
  \right] \;.
 \label{ltensor} 
\end{equation}
Here $p^{\mu} (p^{\prime\mu}$) and $s^{\mu} (s^{\prime\mu}$) 
are the four momentum and four spin of the incoming(outgoing) 
nucleon, respectively. Note that they satisfy the following relations:
\begin{equation}
  p^{\mu}p_{\mu}= p^{\prime\mu}p^{\prime}_{\mu}=M_{p}^{\star 2} \;; \quad
  s^{\mu}s_{\mu}= s^{\prime\mu}s^{\prime}_{\mu}= -1   \;;           \quad
  p^{\mu}s_{\mu}= p^{\prime\mu}s^{\prime}_{\mu}=  0   \;.
 \label{kinem}
\end{equation}

The nuclear-structure information is contained in the scalar response 
function
\begin{equation}
 S({\bf q},\omega) = \sum_{n}
  \left| 
    \langle \Psi_{n} | \hat{\rho}_{S}({\bf q}) | \Psi_{0} \rangle 
  \right|^{2} \delta(\omega-\omega_{n}) \;.
  \label{response}
\end{equation} 
The scalar density
\begin{equation}
  \hat{\rho}_{S}({\bf q}) \equiv
  \int d{\bf x} e^{i{\bf q}\cdot{\bf x}} 
  \bar{\psi}({\bf x}){\psi}({\bf x}) \;,
 \label{rhos}
\end{equation}
is responsible for inducing transitions between the exact nuclear
ground state ($\Psi_{0}$) and an excited state ($\Psi_{n}$) with
excitation energy $\omega_{n}=(E_{n}-E_{0})$. Specific details about 
the calculation of the response are postponed until the next section.

The evaluation of the phase-space and incoming flux factors is
the only task that remains to be performed. The phase-space factor 
is related to the experimentally-detected energy 
$\Big(E'=\sqrt{{\bf p'}^{2}+M^{2}}\Big)$ of the outgoing proton
\begin{equation}
  \int {d{\bf p'} \over (2\pi)^{3}} =
  \int {|{\bf p'}|E' \over (2\pi)^{3}}
        d\Omega \; dE' \;.
 \label{phase}
\end{equation}
The incident flux factor, on the other hand, is evaluated in nuclear 
matter for an incoming nucleon having an effective mass $M_{p}^{\star}$
\begin{equation}
 |{\bf v}| = {|{\bf p}| \over E^{\star}_{\bf p}} \;.
 \label{flux}
\end{equation}
Collecting all appropriate terms leads to the following expression
for the RPWIA spin-dependent cross section 
\begin{equation}
  {d^{2}\sigma_{{\bf s}'{\bf s}} \over d\Omega dE'} =
   {2\pi \over \left(|{\bf p}| / E^{\star}_{{\bf p}}\right)}
   {|{\bf p'}|E' \over (2\pi)^{3}} |t_{S}|^{2} 
   L({\bf p},{\bf s};{\bf p'},{\bf s}') S({\bf q},\omega) \;.
 \label{sigmas2}
\end{equation}

The evaluation of the cross section gets substantially more
complicated when all Lorentz structures of the NN amplitude
are incorporated. The structure of the cross section, however,
remains unchanged:
\begin{equation}
  {d^{2}\sigma_{{\bf s}'{\bf s}} \over d\Omega dE'} =
   {2\pi \over \left(|{\bf p}| / E^{\star}_{{\bf p}}\right)}
   {|{\bf p'}|E' \over (2\pi)^{3}} 
   \sum_{\alpha\beta} t_{\alpha}t_{\beta}^{*}
   L^{\alpha\beta}({\bf p},{\bf s};{\bf p'},{\bf s}')
   S_{\alpha\beta}({\bf q},\omega) \;; \quad 
   (\alpha,\beta={S,V,T,P,A})\;.
 \label{sigmas}
\end{equation}
The main complication arises due to the mixing of the many 
different Lorentz structures of the NN interaction. This
mixing, however, is already familiar from electron-scattering
calculations. Indeed, in electron scattering one must compute
various nuclear-response functions (e.g., mixed vector-tensor
response) due to the anomalous moment of the nucleon. In the 
present case the nuclear-structure information is contained in 
a large set of nuclear-response functions
\begin{equation}
 S_{\alpha\beta}({\bf q},\omega) = \sum_{n}
    \langle 
      \Psi_{n} | \hat{J}_{\alpha}({\bf q}) | \Psi_{0} 
    \rangle 
    \langle 
      \Psi_{n} | \hat{J}_{\beta}({\bf q}) | \Psi_{0} 
    \rangle^{*} \delta(\omega-\omega_{n}) \;,
  \label{sab}
\end{equation}
which are expressed in terms of nuclear currents containing all 
possible Lorentz structures 
\begin{equation}
  \hat{J}^{\alpha}({\bf q}) \equiv
  \int d{\bf x} e^{i{\bf q}\cdot{\bf x}} 
  \bar{\psi}({\bf x})\lambda^{\alpha}{\psi}({\bf x}) \;.
 \label{ja}
\end{equation}
Note that we have introduced the following definitions
\begin{equation}
  \lambda^{\alpha} \equiv
  \Big\{
     1, \gamma^{\mu}, \sigma^{\mu\nu},
    {i\gamma^{5}}, \gamma^{\mu}\gamma^{5} 
  \Big\} \;; \quad
  \bar{\lambda^{\alpha}} \equiv   
  \gamma^{0}\lambda^{\alpha\dagger}\gamma^{0}=\lambda^{\alpha} \;.
 \label{lambdas}
\end{equation}
Likewise, the projectile tensor has been suitable generalized 
from Eq.~(\ref{ltensor}) to
\begin{equation}
  L^{\alpha\beta}({\bf p},{\bf s};{\bf p'},{\bf s}') = 
  {\rm Tr}
  \left[ 
    \lambda^{\alpha}
    \left({ {\rlap/{p}} + M_{p}^{\star} \over 2E^{\star}_{\bf p}  }   \right)
    \left({ 1 + \gamma^{5}{\rlap/{s}} \over 2 }   \right)
    \lambda^{\beta}
    \left({ {\rlap/{p'}} + M_{p}^{\star} \over 2E^{\star}_{\bf p'}  } \right)
    \left({ 1 + \gamma^{5}{\rlap/{s'}} \over 2  } \right)
  \right] \;.
 \label{lij} 
\end{equation}

\subsection{Nuclear response functions}
\label{NRF}

An essential component of the calculation of the cross section
is the nuclear response (as before, we present the relevant 
ideas by considering only the scalar response, for simplicity). 
The nuclear response can be related, in a model-independent way,
to the imaginary part of the scalar polarization~\cite{fetwal71}
\begin{eqnarray}
  S({\bf q},\omega) = &&\sum_{n}
  \left| 
    \langle \Psi_{n} | \hat{\rho}_{S}({\bf q}) | \Psi_{0} \rangle 
  \right|^{2} \delta(\omega-\omega_{n}) \\ \nonumber  =
  {-{1\over\pi}} {\cal I}_{\rm m} &&\sum_{n}
   {\langle \Psi_{0} | \hat{\rho}_{S}(-{\bf q}) | \Psi_{n} \rangle 
    \langle \Psi_{n} | \hat{\rho}_{S}({\bf q})  | \Psi_{0} \rangle 
    \over \omega-\omega_{n}+i\eta} \equiv
  {-{1\over\pi}} {\cal I}_{\rm m} \Pi_{S}({\bf q,\omega}) \;.
 \label{resps}
\end{eqnarray}
This identification is useful because it will enable us to extend 
the calculation beyond the single-particle (or uncorrelated) response. 
Indeed, we will compute all spin-transfer observables by, first, 
calculating the single-particle response of a relativistic mean-field 
ground state and, then, will incorporate long-range correlations 
by solving for the polarization in a relativistic random-phase
approximation (RPA). All nuclear response functions will be calculated 
in infinite nuclear matter. This is an additional and important 
approximation of the model. 

In a mean-field approximation to the nuclear-matter ground state, 
the scalar polarization can be evaluated readily using 
Wick's theorem
\begin{equation}
  i\Pi_{S}(q) = \int {d^{4}k \over (2\pi)^{4}} {\rm Tr}
   \left[
     G(k+q) G(k)
   \right] \;.
 \label{pis}
\end{equation}
Here $G(k)$ is the self-consistent nucleon propagator~\cite{serwal86}
\begin{equation}
   G(k) =  \left({\rlap/{\bar{k}}} + M_{t}^{\star}\right)
     \left[
       {1 \over \bar{k}^{2} - M_{t}^{\star 2} + i\epsilon} +
       {i\pi \over E_{\bf k}^{\star}}
       \delta(\bar{k}^{0}-E_{\bf k}^{\star})
       \theta(k_{F}-|{\bf k}|)
     \right] \;,
 \label{ghartree}
\end{equation}      
written in terms of the Fermi momentum ($k_{F}$). Note that we have
also introduced effective masses and energies for the target nucleons
\begin{equation}
   M_{t}^{\star}=M+\Sigma_{\rm s} \;; \quad
   E^{\star}_{\bf k}=\sqrt{{\bf k}^{2}+M_{t}^{\star 2}} \;; \quad
  \bar{k}^{\mu}=(k^{0}-\Sigma_{\rm v},{\bf k}) \;.
 \label{target}  
\end{equation}
These are shifted from their free-space values by the scalar 
($\Sigma_{\rm s}$) and vector ($\Sigma_{\rm v}$) mean-fields, 
respectively. Also notice that since we are integrating over 
the four-momentum of the target nucleon the contribution from 
the (constant) vector potential can be eliminated by a simple 
change of variables. Formally, then, the response of the 
mean-field ground state is identical to the one of a relativistic 
Fermi gas. All vestige of the relativistic ground-state dynamics 
is subsumed into an effective nucleon mass.

The simplest model of the nuclear response that we will employ is 
a relativistic free Fermi gas. In this model the nuclear response 
consists of the excitation of particle-hole pairs subject to 
the constrains imposed by energy-momentum conservation and the Pauli 
exclusion principle. One can improve this description by taking 
into account, at least at the mean-field level, the interactions 
between the nucleons in the medium. This dressing leads to a shift 
in the nucleon mass but preserves the spectral content of the 
response. Note that since the nuclear response is being probed in 
the spacelike region ($q_{\mu}^{2} <0$) $N\bar{N}$ pairs can not
be excited in these models. They can, however, be virtually
produced. Indeed, the virtual excitation of $N\bar{N}$ pairs
is an important component of the RPA response. RPA correlations play 
a fundamental role in the behavior of isovector observables 
--- particularly in $R_{L}/R_{T}$ --- and their inclusion is one of 
the central results from the present paper.

We now generalize the above expression for the scalar polarization
to an arbitrary Lorentz structure 
\begin{equation}
   i\Pi^{\alpha\beta}_{ab}(q) = 
     \int {d^{4}k \over (2\pi)^{4}} {\rm Tr}
     \left[
       \lambda^{\alpha}\tau_{a}G(k+q) 
       \lambda^{\beta }\tau_{b}G(k) 
     \right] \;.
 \label{piab}
\end{equation}
Note that the appropriate isospin matrices have been included 
to reflect the isovector character of the $(\vec{p},\vec{n})$
reaction. Many-body correlations are incorporated by considering 
the residual interaction between the particle and the hole. 
Formally, this is accomplished by solving Dyson's equation for 
the correlated RPA propagator~\cite{fetwal71}. The RPA equation 
is characterized by an infinite summation of the lowest-order 
(uncorrelated) polarization
\begin{equation}
   \widetilde{\Pi}^{\alpha\beta}_{ab} = 
             {\Pi}^{\alpha\beta}_{ab} +
             {\Pi}^{\alpha 5}_{ac} 
             V^{(\pi)}_{cd}
             {\Pi}^{5  \beta}_{db}    +
             {\Pi}^{\alpha \mu}_{ac} 
             V^{(\rho)}_{\mu\nu;cd}
             {\Pi}^{\nu  \beta}_{db}  + \ldots \;.
  \label{pirpa}
\end{equation}
Note that we have written the residual particle-hole interaction 
in terms of $\pi + \rho$ contributions (the issue of short-range 
correlations is postponed until the next section)
\begin{eqnarray}
  V^{(\pi)}_{ab}(q)                   &=&
      \delta_{ab} V^{(\pi)}(q)         = 
      \delta_{ab} f_{\pi}^{2}\Delta(q) =
      \delta_{ab}{f_{\pi}^{2} \over q_{\mu}^{2} - m_{\pi}^{2}} 
      \;, \\ \nonumber
  V^{(\rho)}_{\mu\nu;ab}(q)                &=& 
      \delta_{ab} V^{(\rho)}_{\mu\nu}(q)    = 
      \delta_{ab} g_{\rho}^{2}D_{\mu\nu}(q) =
      \delta_{ab} 
      \left[-g_{\mu\nu} + {q_{\mu}q_{\nu}\over m_{\rho}^{2}}\right]
      {g_{\rho}^{2} \over q_{\mu}^{2} - m_{\rho}^{2}} \;.
 \label{vpirho}
\end{eqnarray}
Because the ground-state of nuclear matter is assumed to be isospin 
symmetric, and because both ($\pi$ and $\rho$) propagators are diagonal 
in isospin, the isospin structure of the polarizations is simple and 
given by
\begin{equation}
   \widetilde{\Pi}^{\alpha\beta}_{ab}(q) = \delta_{ab}
   \widetilde{\Pi}^{\alpha\beta}(q) \;; \quad
             {\Pi}^{\alpha\beta}_{ab}(q) = \delta_{ab}
             {\Pi}^{\alpha\beta}(q) \;,
 \label{ispin}
\end{equation}
where the RPA polarizations satisfy 
\begin{equation}
   \widetilde{\Pi}^{\alpha\beta}(q)  =
             {\Pi}^{\alpha\beta}(q)  +
             {\Pi}^{\alpha i}(q) 
             \widetilde{V}_{ij}(q) 
             {\Pi}^{j \beta}(q)      \;.
 \label{piabrpa}
\end{equation}
The RPA polarizations $(\widetilde{\Pi})$ have been written in terms 
of the lowest order polarizations and the medium-modified 
isovector interaction $(\widetilde{V})$. The latin indices
$i$ and $j$, with values in the range $-1,0,1,2,3$, are used to 
denote the ``elementary'' coupling of the nucleon to, either, the 
$\pi(i=-1)$ or the $\rho(i=0,1,2,3)$ mesons (this coupling is
represented by a dot in Fig.~\ref{figone}). In contrast, the greek indices 
$\alpha$ and $\beta$ represent, as before, any of the five Lorentz 
structures ($S,V,T,P,A$) present in the NN amplitude (we have represented
this coupling by a cross in Fig.~\ref{figone}). The RPA dynamics is, thus, 
fully contained in the medium-modified isovector interaction satisfying 
the following Dyson's equation:
\begin{equation}
   \widetilde{V}_{ij}(q)             =
             {V}_{ij}(q)             +
             {V}_{ik}(q)
             {\Pi}^{kl}(q)
             \widetilde{V}_{lj}(q) \;.
 \label{vijrpa}
\end{equation}
Note that the free-space interaction $V_{ij}$ is diagonal 
\begin{equation}
  V_{ij} = \left(
   \begin{array}{cc}
            V^{(\pi)}     &            0         \\
                0         &  V^{(\rho)}_{\mu\nu} 
  \end{array} \right) \;,
 \label{vij}
\end{equation}
and that the elementary $NN-$meson vertex (denoted by
$\Lambda^{i}$ in the next section) will be dictated by 
our choice of residual particle-hole interaction.

\subsection{Residual particle-hole interaction: $\pi + \rho + g'$}
\label{pirho}

The residual interaction consists of $\pi + \rho + g'$ contributions. 
For the $NN\pi$ coupling we assume a pseudovector representation. It 
is convenient to adopt this, as opposed to a pseudoscalar, representation 
because it incorporates the correct low-energy pion dynamics without 
sensitive cancellations~\cite{serwal86}. The $\rho-$meson contains 
a vector as well as a tensor coupling to the nucleon. With these choices 
we have specified completely the elementary $NN\pi$ and $NN\rho$ vertices
to be used in Eqs.~(\ref{piabrpa})~and~(\ref{vijrpa}):
\begin{equation}
  \Lambda^{i} = \cases{   
     \displaystyle{{\rlap/{q} \over m_{\pi}}\gamma^{5}} 
      & if $i=\pi$; \cr
     \displaystyle{\left(
       \gamma^{\mu}\pm iC_{\rho}\sigma_{\mu\nu}{q_{\nu} \over 2M}
                   \right)}
      & if $i=\rho$. \cr}
 \label{lambdai}
\end{equation}
Note that the plus(minus) sign should be used when the four-momentum
of the $\rho-$meson flows into(away from) the vertex (in the case of
the pion the minus sign has been incorporated into the definition of 
$V^{(\pi)}$). The $\pi$ and $\rho$ components of the NN potential in 
the nonrelativistic limit are well known. The one-pion-exchange
contribution is given by
\begin{equation}
   V_{\pi}({\bf q})= -{f_{\pi}^{2} \over m_{\pi}^{2}}
                      {({\bf \sigma}_{1}\cdot{\bf q})
                       ({\bf \sigma}_{2}\cdot{\bf q}) \over
                       {\bf q}^{2} + m_{\pi}^{2}}
                      ({\bf \tau}_{1}\cdot{\bf \tau}_{2}) \;,
 \label{vpion}
\end{equation}
where the ``spin-longitudinal'' coupling
\begin{equation}
    {f_{\pi}^{2} \over 4\pi} = 
    \left( m_{\pi} \over 2M \right)^{2}
    {g_{\pi}^{2} \over 4\pi}  \;,
 \label{ftogpi}
\end{equation}
has been defined in terms of the couplings and masses listed
in Table~\ref{tableone}. Although the $\rho$-meson contains a 
vector as well as a tensor coupling to the nucleon, the $NN\rho$ 
coupling is dominated by the large tensor contribution which,
in particular, is responsible for generating the transverse character 
of the interaction
\begin{equation}
   V_{\rho}({\bf q})= -{f_{\rho}^{2} \over m_{\rho}^{2}}
                      {({\bf \sigma}_{1}\times{\bf q}) \cdot
                       ({\bf \sigma}_{2}\times{\bf q}) \over
                       {\bf q}^{2} + m_{\rho}^{2}}
                      ({\bf \tau}_{1}\cdot{\bf \tau}_{2}) \;.
 \label{vrho}
\end{equation}
Note that we have introduced the ``spin-transverse'' coupling
\begin{equation}
    {f_{\rho}^{2} \over 4\pi} = 
    \left( m_{\rho} \over 2M \right)^{2}
    {g_{\rho}^{2} \over 4\pi}C_{\rho}^{2}  \;,
 \label{ftogrho}
\end{equation}
in terms of the (large) tensor-to-vector ratio $C_{\rho}$.
As it stands, the interaction is extremely attractive in the spin-spin 
channel. In order to regularize the large spin-spin component of the 
interaction one simulates the effect from repulsive short-range 
correlations by introducing a phenomenological Landau-Migdal 
parameter $g'$
\begin{equation}
   V_{g'}({\bf q})= {f_{\pi}^{2} \over m_{\pi}^{2}} g'
                    ({\bf \sigma}_{1}\cdot{\bf \sigma}_{2})
                    ({\bf \tau}_{1}\cdot{\bf \tau}_{2}) \;.
 \label{vgprime}
\end{equation}
The effect of $g'$ will be incorporated by modifying, separately,
the $\pi$ and (the transverse component of) the $\rho$ propagators. 
This can be accomplished with the use of the following identity:
\begin{equation}
 ({\bf \sigma}_{1}\cdot{\bf \sigma}_{2})=
 \Big[
  ({\bf \sigma}_{1}\cdot{\bf q})({\bf \sigma}_{2}\cdot{\bf q}) +
  ({\bf \sigma}_{1}\times{\bf q})\cdot({\bf \sigma}_{2}\times{\bf q})
 \Big] \;.
 \label{ident}
\end{equation}
In this way, the spin-longitudinal and spin-transverse components of  
$V_{g'}$ get absorbed into redefinitions of ``effective'' $\pi$ and 
$\rho$ propagators
\begin{eqnarray}
  \Delta_{\pi}(q_{\mu}^{2}) &=& 
     {1 \over q_{\mu}^{2} - m_{\pi}^{2}} \rightarrow 
     \Big[
      {1 \over q_{\mu}^{2} - m_{\pi}^{2}} - 
      {g'_{\pi} \over q_{\mu}^{2}} 
     \Big] \;, \\ \nonumber
  \Delta_{\rho}(q_{\mu}^{2}) &=& 
     {1 \over q_{\mu}^{2} - m_{\rho}^{2}} \rightarrow 
     \Big[
      {1 \over q_{\mu}^{2} - m_{\rho}^{2}} - 
      {g'_{\rho} \over q_{\mu}^{2}} 
     \Big] \;,
 \label{effprop}
\end{eqnarray}
where
\begin{equation}
  g'_{\pi}  \equiv g' \;; \quad
  g'_{\rho} \equiv g'
   \left[
     \left({f_{\pi}^{2}  / m_{\pi}^{2}} \right)  \Big/
     \left({f_{\rho}^{2} / m_{\rho}^{2}}\right) 
   \right] \;.
 \label{gprimes}
\end{equation}
In all that follows we will assume the standard value of $g'=0.7$
for the phenomenological Landau-Migdal parameter. Note that in this 
case $g_{\pi}'=0.7$ and $g_{\rho}'\sim 0.65$ for the parameters 
given in Table~\ref{tableone}. We have displayed the resulting 
isovector interaction in Fig.~\ref{figtwo}.

We now examine isoscalar effects, from a reduced nucleon mass in the 
medium, on the isovector response. At the simplest level a reduction 
in the value of the nucleon mass leads to a shift in the position and 
to an increase in the width of the quasielastic peak, i.e.,
\begin{equation}
  \omega_{QE} = {Q^{2} \over 2M} \rightarrow  
                {Q^{2} \over 2M^{\star}} \;; \quad
  \Delta\omega \simeq {qk_{F} \over M} \rightarrow 
                      {qk_{F} \over M^{\star}} \;.
 \label{qepeak}
\end{equation}
This simple realization, however, has nontrivial consequences 
for the case of the transverse response. Relative to a free
Fermi-gas calculation the transverse response measured in
quasielastic electron scattering appears to be quenched and 
hardened. This observation provides strong evidence in favor 
of strong repulsive correlations in the transverse spin-isospin 
channel. Note, however, that the uncorrelated response of 
a relativistic mean-field ground state is already ``hardened'' 
relative to the Fermi gas response. This is a simple consequence 
of the in-medium reduction of the nucleon mass. Moreover, the 
transverse $(e,e')$ response is dominated by the large isovector anomalous 
moment of the nucleon. Thus, the integrated response is insensitive 
to a change in the value of the nucleon mass. Since the width increases 
as $M^{\star}$ is reduced, the distribution appears to be quenched 
relative to the Fermi-gas response. Hence, the uncorrelated response 
of a relativistic mean-field ground state accounts for most of the 
features --- quenching and hardening --- observed experimentally.
Indeed, in Fig.~\ref{figthree} a comparison is made between 
relativistic mean-field calculations of the transverse response 
and experimental data for ${}^{40}$Ca$(e,e')$ at momentum transfers 
of $q=410$~and~$q=550$~MeV~\cite{barr83}. The dotted line shows the 
results from a finite-nucleus calculation of the Hartree (uncorrelated) 
response. Good agreement with experiment is found for the low-energy 
side of the quasielastic peak. The underestimation of transverse 
strength on the high-energy side of the peak, believed to be dominated 
by isobar formation and meson-exchange currents, is a common 
shortcoming of most ``one-nucleon'' models. The fact that experiment 
shows no hardening of the transverse response relative to the 
Hartree predictions is one of the important results of this 
comparison. 

In Fig.~\ref{figthree} we have also included relativistic
calculation of the nuclear-matter response with (solid line) and without 
(dashed line) RPA correlations. Based on our finite-nucleus results we have 
adjusted the ``spin-transverse'' component of the Landau-Migdal parameter 
to minimize the effect from RPA correlations. This could be achieved 
by selecting $g_{\rho}^{'}$ in the range $g_{\rho}^{'}\sim (0.3-0.4)$. 
This represents a substantial reduction from its conventional $M^{*}=M$ 
value of $g_{\rho}^{'}\sim 0.65$. We will use 
$g_{\rho}^{'} \equiv 0.3$ for $M^{\star}\ne M$ and 
$g_{\rho}^{'} \equiv 0.65$ for $M^{\star}=M$, in all that follows.
In Fig.~\ref{figfour} we have displayed the residual isovector interaction 
for the relativistic $M^{\star} \ne M$ case. It is important to realize 
that the value of $g_{\rho}^{'}$ is regarded as a purely phenomenological 
parameter constrained by electron-scattering data. 

The longitudinal component of the residual interaction is also
sensitive to a reduced effective nucleon mass in the medium.
However, in contrast to the transverse component this modification 
does not emerge from a ``phenomenological'' tuning of parameters.
Rather, it is a genuine dynamical effect that reduces the effective
$NN\pi$ coupling in the medium. The origin of this reduction is 
as follows. In the medium, the pion-mediated NN interaction 
is given, according to Eq.~(\ref{vijrpa}), by
\begin{eqnarray}
   \widetilde{V}^{(\pi)}(q)           &=&
      {\epsilon}_{\pi}^{-1}(q;k_{F})             
                {V}^{(\pi)}(q) \;, \\ \nonumber 
      {\epsilon}_{\pi}(q;k_{F})   &\equiv&
      1 - {V}^{(\pi)}(q)\Pi^{PV}(q) \;,
 \label{vrpapion}
\end{eqnarray}
where we have introduced the pion dimesic function ${\epsilon}_{\pi}$
and have defined the pseudovector polarization by~\cite{dawpie91}
\begin{equation}
   i\Pi^{PV}(q) = 
     \left( {q_{\mu} \over m_{\pi}} \right)
     \left( {q_{\nu} \over m_{\pi}} \right)
     \int {d^{4}k \over (2\pi)^{4}} {\rm Tr}
     \left[
       \gamma^{\mu}\gamma^{5}G(k+q)
       \gamma^{\nu}\gamma^{5}G(k)
     \right] \;.
 \label{pipvpion}
\end{equation}
In the absence of a mass term from Walecka's mean-field Lagrangian, 
the axial-vector current would be conserved and the pseudovector 
polarization would vanish. Thus, any finite contribution to $\Pi^{PV}$ 
must arise --- and be proportional to --- the nucleon mass. Indeed,
in a previous study we determined the following behavior for
the pion dimesic function in the static limit:
\begin{equation}
   {\epsilon}_{\pi}(|{\bf q}|;k_{F}) =
    1- {{\bf q}^{2} \over {\bf q}^{2} + m_{\pi}^{2}}
       f_{\pi}^{2} M^{\star 2} {2 \over \pi^{2} |{\bf q}|}
       \int_{0}^{k_{F}} dk {k \over E^{\star}_{\bf k}}
       \ln \left|
         {|{\bf q}| + 2k \over |{\bf q}| - 2k}
       \right| \;,
 \label{pidimes}
\end{equation}
which suggests the following limits for the effective $NN\pi$ coupling 
in the medium
\begin{equation}
    {f_{\pi}^{\star 2} \over f_{\pi}^{2}} =
     \cases{ 
       \left({M^{\star} \over M}\right)
        & at low density; \cr
       \left({M^{\star} \over M}\right)^{2}
        & at high density. \cr}
 \label{fpistar}
\end{equation}
Note that the effective coupling is strongly density dependent.
Indeed, the suppression of the effective $NN\pi$ coupling with
increasing density more than compensates for the increase in the 
value of the integral leading, in particular, to no pion 
condensation --- even in the absence of a phenomenological 
Landau-Migdal parameter ($g'\equiv 0$)~\cite{dawpie91}. This 
dynamical suppression of the $NN\pi$ coupling in the medium is 
instrumental in reducing the enhancement of the longitudinal 
spin-response relative to nonrelativistic ($M^{\star}=M$) 
predictions~\cite{horpie93}.

\subsection{Spin-transfer observables}
\label{sto}

Spin-transfer observables for the $(\vec{p},\vec{n})$ 
[and also for the $(\vec{p},\vec{p}')$] reaction
can be obtained as linear combinations of the spin-dependent 
cross sections defined in Eq.~(\ref{sigmas}). In particular, 
the out-of-plane observables $({\bf s}={\bf s}'=\hat{\bf n})$ 
are given by
%\begin{mathletters}
\begin{eqnarray}
  \sigma &=&
  \Big( 
      \sigma_{++} + \sigma_{+-} 
    + \sigma_{-+} + \sigma_{--} 
  \Big)  \;, \\
  \sigma P &=&
  \Big( 
      \sigma_{++} + \sigma_{+-} 
    - \sigma_{-+} - \sigma_{--} 
  \Big)  \;, \\
  \sigma A_{y} &=&
  \Big( 
      \sigma_{++} - \sigma_{+-} 
    + \sigma_{-+} - \sigma_{--} 
  \Big)  \;, \\
  \sigma D_{NN} &=&
  \Big( 
      \sigma_{++} - \sigma_{+-} 
    - \sigma_{-+} + \sigma_{--} 
  \Big)  \;,
 \label{outplane}
\end{eqnarray}
%\end{mathletters}
where we have introduced the following simplified notation
\begin{eqnarray*} 
   {d^{2}\sigma_{\pm\hat{n}\pm\hat{n}} \over d\Omega dE'}
   \equiv \sigma_{\pm\pm} \;. 
\end{eqnarray*}     
Since the conservation of parity forces sideways and longitudinal 
polarizations and analyzing powers to vanish, the remaining four 
independent observables, namely, $D_{S'S}, D_{L'S}, D_{S'L},$ 
and $D_{L'L}$ can all be obtained from the in-plane cross sections 
in analogy to $D_{NN}$. 

	The simplicity of the reaction mechanism (i.e., quasifree 
knockout) in quasielastic $(\vec{p},\vec{n})$ scattering makes the 
study of the nuclear spin-isospin response particularly clean.
That the dominant mechanism in the reaction is, indeed, quasifree 
knockout can be justified by studying certain combinations of spin 
observables. Specifically, we consider
$(P-A_{y})$ and $[(D_{S'L}+D_{L'S}) -
(D_{L'L}+D_{S'S})\tan(\theta_{\rm lab})]$.
Due to time-reversal invariance these combinations are identically 
zero in elastic scattering and have been plotted in Fig.~\ref{figfive} 
for ${}^{40}$Ca$(\vec{p},\vec{n})$ at $q=1.72$~fm$^{-1}$. The fact that 
both of these observables are consistent with zero all across the
quasielastic region is strong evidence in support of a quasifree 
mechanism for the reaction.

One of the most appealing features of having a complete set 
of spin-transfer observables is the possibility of extracting 
nuclear-response functions, such as the spin-longitudinal response, 
which are not accessible with electromagnetic probes. Under certain 
approximations, these responses can be directly related to linear 
combinations of the standard spin-transfer 
coefficients~\cite{carey84,blesz82,chan90} 
   \begin{eqnarray}
       D_{0}&=&{1 \over 4} 
        \left[ 
          (1+D_{NN})+
          (D_{S'S}+D_{L'L})\cos(\bar{\theta})-
          (D_{S'L}-D_{L'S})\sin(\bar{\theta})
        \right] \;, 
       \label{d0} \\
       D_{N}&=&{1 \over 4} 
        \left[ 
          (1+D_{NN})-
          (D_{S'S}+D_{L'L})\cos(\bar{\theta})+
          (D_{S'L}-D_{L'S})\sin(\bar{\theta})
        \right] \;, 
       \label{dn} \\
       D_{L}&=&{1 \over 4} 
        \left[ 
          (1-D_{NN})+
          (D_{S'S}-D_{L'L})\;{\rm sec}({\theta_{\rm lab}})
        \right] \;, 
       \label{dl} \\
       D_{T}&=&{1 \over 4} 
        \left[ 
          (1-D_{NN})-
          (D_{S'S}-D_{L'L})\;{\rm sec}({\theta_{\rm lab}})
        \right] \;, 
       \label{dt} 
   \end{eqnarray}
where we have defined 
$\bar{\theta}\equiv{\theta_{\rm cm}}-\theta_{\rm lab}$.
Note, in particular, that this new set of polarization-transfer 
observables satisfy the constraint
\begin{equation}
  D_{0}+D_{N}+D_{L}+D_{T}=1 \;.
 \label{dsum}
\end{equation}
Nuclear responses per target nucleon, $R_{i}({\bf q},\omega)$, 
can now be defined according to
\begin{equation}
  R_{i}({\bf q},\omega) A_{\rm eff} \equiv
 \left[
    {1 \over \sigma_{NN}}
    \left({d^{2}\sigma \over d\Omega dE'}\right)
 \right]
 \left[
    {\big(D_{i}\big)_{NA} \over \big(D_{i}\big)_{NN}}
 \right] \;,
 \label{defresp}
\end{equation}
where $A_{\rm eff}$ represents the effective number of
nucleons participating in the reaction.

Before proceeding further we show evidence in support of 
the above definition of the response. In Fig.~\ref{figsix}
we have calculated spin-longitudinal and spin-transverse
responses from nuclear matter using two different procedures.
In one of them, the responses are reconstructed from the
spin-transfer observables as outlined above. In the second
method the responses are computed by assuming that only a 
pion (for $R_{L}$) or a $\rho$-meson (for $R_{T}$) couple
to the target nucleons. Thus, for $R_{L}$ we compute the 
response directly from the pseudovector polarization given 
in Eq.~(\ref{pipvpion}). For the transverse response we simply 
report the electron-scattering result since it is known to 
be dominated by the anomalous (isovector-tensor) moment
of the nucleon. The agreement between the two procedures,
for, both, Fermi gas and RPA responses, lends credibility 
to the approach. In particular we note, as previously advertised, 
that the longitudinal response becomes soften and enhanced, while
the transverse hardened and quenched, relative to Fermi gas 
predictions. We now examine how relativistic effects, mainly 
from a reduced nucleon mass, modify these findings.

\section{Results}
\label{secres}

We now proceed to show RPWIA results for all spin-transfer
observables. Results will be presented using four different
approximations. The simplest approximation consists of treating
the nucleus as a relativistic free Fermi gas. In this case the 
various nuclear responses arise from imposing simple constraints 
such as energy-momentum conservation and the Pauli principle. 
Thus, no important deviations in the values of spin observables 
are expected since, neither, the underlying NN interaction nor 
the collective response of the target are modified. Next we consider 
the uncorrelated response of a relativistic Hartree ground state. In 
this case the propagation of a nucleon through the medium is modified 
by its interaction with the mean-field. This results in a reduction 
of the nucleon mass which now modifies the underlying NN interaction 
since matrix elements of the amplitude are being computed with in-medium 
(as opposed to free) spinors. The last two approximations are obtained 
from the previous two by incorporating RPA correlations into the nuclear
 response. In both cases the longitudinal component of the residual 
interaction is the same while the transverse component is constrained 
by the transverse response measured in quasielastic electron scattering, 
as previously discussed. Relevant parameters characterizing the residual
interaction and the effective nucleon masses are given in 
Table~\ref{tableone} and Table~\ref{tabletwo}, respectively. Finally, 
we have used the FA90 phase-shift solution of Arndt~\cite{arndt} 
to generate the relativistic parameterization of the NN amplitude.

In Fig.~\ref{figseven} and Fig.~\ref{figeight} the complete 
set of spin observables at $q=1.72$~fm$^{-1}$ is compared to 
the experimental data. The differential cross section has 
been divided by the single-nucleon value and reported as a
spin-averaged response 
\begin{equation}
  R({\bf q},\omega) =
   \left[
    {1 \over \sigma_{NN}}
    \left({d^{2}\sigma \over d\Omega dE'}\right)
   \right] \;.
 \label{rqw}
\end{equation}
The experimental value for the single-nucleon cross section 
was obtained by integrating the strength under the quasifree 
peak for the ${}^{2}$H$(\vec{p},\vec{n})$ reaction and was
reported to be 11.5 mb/sr~\cite{chen93}. This agrees
well the free value of 11.6 mb/sr~\cite{arndt} and also with 
our own value of 11.4 mb/sr obtained from a ``nuclear-matter'' 
calculation in the limit of $k_{F}\rightarrow 0$. Note, we have
divided our theoretical cross sections by 10.9 mb/sr which is 
the appropriate value at fixed $q=1.72$~fm$^{-1}$ (as opposed to
fixed angle). Finally, we have adopted the value of $A_{\rm eff}=6$ 
for the effective number of nucleons participating in the reaction and 
have shifted all observables by the reaction Q-value assumed to be 
$Q=18.1$~MeV~\cite{hormur88}. 

The Fermi-gas response (dotted line)
peaks at an energy loss corresponding to free NN scattering, namely
$\omega_{QE} \simeq (\sqrt{{\bf q}^{2}+M^{2}} - M) + Q \simeq 77$~MeV. 
Moreover, the integrated Fermi-gas strength equals 5.9. We attribute 
the small difference between this value and $A_{\rm eff}$ to Pauli 
blocking. Since the momentum transfer ($q=1.72$~fm$^{-1}$) is slightly 
smaller than twice the Fermi momentum ($2k_{F}=1.88$~fm$^{-1}$) a few 
transitions are Pauli blocked as is evident in the behavior of the 
response on the low$-\omega$ side of the peak. Thus, aside from a 
small correction due to Pauli blocking, the Fermi-gas response 
consists of a simple redistribution of single-particle strength.

The Hartree ($M^{\star} \ne M$) response is depicted by the
dot-dashed line (almost indistinguishable from the solid line).
The shift in the position, and the increase 
in the width, of the quasielastic peak relative to the 
Fermi-gas response are clearly evident and supported by experiment. 
Note, however, that the strength under the Hartree peak amounts to 
only 4.9. This reduction, which now underestimates the data, is 
caused by a modification of the effective NN interaction in the
medium since $\bar{{\cal U}}(M^{\star}){\rlap/{q} \gamma^{5}}
{\cal U}(M^{\star})$ is less than $\bar{{\cal U}}(M){\rlap/{q} \gamma^{5}} 
{\cal U}(M)$.

The RPA response of the Fermi-gas ground state (dashed line)
is hardened and quenched relative to the Fermi-gas 
response. Note that the integrated RPA strength is 5.5. This 
suggests that, at least for the spin-averaged response, the 
repulsive character of the transverse interaction dominates over 
the attractive longitudinal component (see Fig~.\ref{figtwo}). 
In contrast, RPA correlations have no observable effect on the
$M^{\star} \ne M$ spin-averaged response (solid line). At this 
momentum transfer, the weaker repulsion from the transverse 
channel is almost completely cancelled by the longitudinal 
attraction (see Fig.~\ref{figfour}) resulting in an integrated 
response of 4.9, as in the Hartree case.

As for the remaining spin observables, some systematic trends 
emerge. First, RPA correlations with free masses (dashed lines) 
generate dramatic changes with respect to the Fermi-gas values 
(dotted lines) and give a poor description of the data. In contrast, 
relativistic ($M^{\star} \ne M$) RPA correlations (solid lines) lead 
to a good description of the data and, in all cases, to an improvement 
over the Fermi gas predictions. Perhaps the only case in which the 
agreement is not as good is for $D_{NN}$. Notice, however, that this 
discrepancy is present even at the level of the free NN observables
(see Table~\ref{tablethree}). This suggests that a more faithful
representation of the many-body dynamics could be obtained by 
removing the ``single-nucleon'' component of the observable, as
was done for the cross section.

In Fig.~\ref{fignine} we have plotted the new set of 
polarization-transfer observables relative to their free NN 
values. The fact that the spin-independent component $D_{0}$ 
is ill determined is a reflection of the spin-dependent character
of the $(\vec{p},\vec{n})$ reaction. This, however, does not pose
any serious limitation on the analysis since only three of the four 
observables are known to be independent [see Eq.~(\ref{dsum})].
The data shows, if at all, very small deviations from unity.
The Fermi gas (dotted line), and to a lesser extend the Hartree 
(dot-dashed line), results are also close to unity but, arguably, the 
best overall description of the data is obtained with the relativistic 
RPA calculation (solid line). Instead, a poor description of the 
experimental observables is obtained whenever RPA correlations with
free nucleon masses are included (dashed line). Yet, these results 
conform to the notion of a hardened transverse ($D_{N}$ and $D_{T}$) 
and a softened longitudinal ($D_{L}$) response. Note, the actual 
responses are obtained from the above observables by multiplying them 
by the spin-averaged response $R({\bf q},\omega)$ 
(see Fig.~\ref{figseven}). Perhaps the most prominent feature of 
our results is the mild enhancement predicted for $D_{L}$ by the 
relativistic RPA calculation (solid line). This behavior is a direct 
consequence of the dynamical suppression of the $NN\pi$ coupling in 
the medium, as discussed in Sec.~\ref{pirho}.

The spin-longitudinal to spin-transverse ratio, $R_{L}/R_{T}$, 
is shown in Fig.~\ref{figten}. This plot summarizes --- and
dramatizes --- some of the findings of the previous plot. 
Fermi-gas predictions (dotted line) are seen to be consistent with 
unity while Hartree results (dot-dashed line) show a mild suppression 
arising from a modified NN interaction in the medium. RPA correlations 
with free nucleon masses (dashed line) suggest a large enhancement 
in the ratio which is not observed experimentally and is reminiscent of 
the original nonrelativistic predictions. Finally, dynamical effects from 
a reduced nucleon mass in the medium generate a large suppression
in the ratio (solid line) relative to the $M^{\star}=M$ predictions,
in good agreement with experiment.

One important test for all models of the $(\vec{p},\vec{n})$ reaction
is the momentum-transfer dependence of the ratio. This is so because 
competing models predict a different behavior with momentum transfer
of the residual interaction. Specifically, a residual interaction 
having a transverse component modified by an in-medium reduction of 
the $\rho$-meson mass~\cite{brown90} gives a description of $R_{L}/R_{T}$ 
at $q\simeq 1.72$~fm$^{-1}$ of similar quality to the relativistic 
case~\cite{browam94}. Yet, these two models predict a vastly different 
momentum-transfer dependence for the ratio. In particular, a reduced
$\rho$-meson mass generates an enhancement --- rather than a 
quenching --- in $R_{L}/R_{T}$ at $q\simeq 2.5$~fm$^{-1}$. This 
arises from a transverse component that has become attractive 
at {\hbox{$q{\lower.40ex\hbox{$>$}\atop\raise.20ex\hbox{$\sim$}}$}}
2.2 fm$^{-1}$ due to a faster falloff induced by the in-medium 
reduction of the $\rho-$meson mass. In contrast, in the relativistic 
model the qualitative features of the isovector interaction, namely, 
attractive longitudinal and repulsive transverse components, remain 
unchanged over the entire range of momentum transfers sampled in the 
experiment~\cite{taddeu94}. Note, these two models also make definite, 
and most likely different, predictions for the momentum-transfer 
dependence of the transverse response measured in inclusive electron 
scattering.

In Fig.~\ref{figeleven} we display the relativistic predictions
for the momentum-transfer dependence of $R_{L}/R_{T}$ in ${}^{12}$C.
Since the A dependence of the spin-observables is known to be weak, 
the good agreement between theory and experiment at $q=1.72$~fm$^{-1}$ 
is not surprising. One also observes that the trends inferred from the 
$q=1.72$~fm$^{-1}$ calculations are preserved at low-momentum 
transfer, namely, a Fermi-gas value close to unity (dotted line), a 
slight suppression in the Hartree result (dot-dashed line), and 
a mild enhancement at low$-\omega$ in the RPA value (solid line) which,
however, now underestimates the experimental ratio (note, for a 
preliminary experimental report see Ref.~\cite{taddeu94}). In the 
absence of high$-q$ data one could resolve this small discrepancy with a 
fine tuning of parameters. Indeed, a slight increase in $g_{\rho}^{'}$, 
or, alternatively, a slight decrease in $g_{\pi}^{'}$, could enhance 
the ratio at low$-\omega$ and, thus, bring the calculations into 
agreement with experiment. Explaining the reported quenching of 
$R_{L}/R_{T}$ at $q=2.5$~fm$^{-1}$~\cite{taddeu94}, however, is likely 
to require physics that is absent from our model. Note that at 
$q=2.5$~fm$^{-1}$ even the Fermi-gas (dotted line) and Hartree 
(dot-dashed line) ratios are already enhanced at low$-\omega$ relative 
to the free NN ratio. Moreover, since the transverse component of the 
interaction is never as attractive as the longitudinal one (at least 
within the range plotted in Fig.~\ref{figfour}) the RPA ratio will exceed 
the Hartree value --- and, thus, unity --- at all values of $q$.

Some insight into the missing physics can be obtained from the 
distorted-wave calculations of Ichimura and collaborators. 
For the original $q=1.72$~fm$^{-1}$ calculation, it was shown 
that while distortions provide an overall reduction of the cross 
section, they do so without substantially modifying the distribution 
of strength~\cite{ichim89}. Recently, however, they have suggested 
that distortions play a significant role in $R_{L}/R_{T}$ at, both, 
$q=1.2$~fm$^{-1}$ and $q=2.5$~fm$^{-1}$~\cite{taddeu94}. In particular, 
they found that, relative to the free NN values, distortions enhance 
the ratio at $q=1.2$~fm$^{-1}$ but quench it at $q=2.5$~fm$^{-1}$. 
Therefore, it is conceivable that after the inclusion of distortions 
relativistic RPA calculations could yield a reasonable description of 
the experimental ratio for all three values of $q$. This expectation 
is currently under investigation.

\section{Conclusions}
\label{secconcl}

	We have calculated all spin-transfer observables for the
quasielastic $(\vec{p},\vec{n})$ reaction using a relativistic 
random-phase approximation to the Walecka model. Isoscalar effects 
arising from a dynamical reduction in the nucleon mass are responsible 
for a shift in the position and for an increase in the width of the 
quasielastic peak. These two features, by themselves and without
RPA correlations, are sufficient to explain the ``quenching'' and 
``hardening'' of the transverse response. Moreover, the reduced value 
of the nucleon mass generates a dynamical suppression of the $NN\pi$ 
coupling in the medium. This effect reduces the enhancement of the 
longitudinal response relative to an equivalent ``nonrelativistic'' 
($M^{\star}\equiv M$) calculation. As a consequence, a good description 
of all spin-transfer observables is obtained at $q=1.72$~fm$^{-1}$.
In particular, we observed no enhancement in the spin-longitudinal 
to spin-transverse ratio, in agreement with experiment.

Brown and Wambach have offered an alternative explanation for the 
lack of an enhancement in $R_{L}/R_{T}$ at $q=1.72$~fm$^{-1}$ by
invoking a rescaling of the $\rho$-meson mass in the nuclear 
medium~\cite{browam94}. However, a recent report on the measurement 
of $R_{L}/R_{T}$ at $q=1.2$~fm$^{-1}$ and $q=2.5$~fm$^{-1}$ suggests 
that the real picture might be more complicated than the one presented 
by either of these two models~\cite{taddeu94}. Particularly noteworthy 
are the results at $q=2.5$~fm$^{-1}$. The experimental results seem to 
confirm the suppression at low-energy loss of $R_{L}/R_{T}$ predicted 
by the ($m_{\rho}^{\star}$) rescaling model. Yet, the data does not 
support the rapid variation with energy loss suggested by the model. 
Specifically, the rescaling model predicts $R_{L}/R_{T}\sim 1$ at the 
position of the quasielastic peak while the data remains constant at 
$R_{L}/R_{T}\sim 0.6$. For reference, the relativistic model overpredicts 
the ratio over the whole low$-\omega$ region of the quasielastic peak.

The distorted-wave calculations of Ichimura and collaborators 
might shed some light into the problem. We should note that these 
calculations do not incorporate any ``novel'' effect so they do 
overpredict the spin-longitudinal to spin-transverse ratio in RPA. 
Yet, their realistic treatment of distortions is very valuable. For 
example, these calculations revealed a modest effect of distortions
on $R_{L}/R_{T}$ at $q=1.72$~fm$^{-1}$~\cite{ichim89}. A recent report 
suggests, however, that distortions are important in enhancing and 
quenching the ratio at $q=1.2$~fm$^{-1}$ and $q=2.5$~fm$^{-1}$, 
respectively~\cite{taddeu94}. Indeed, the distorted-wave calculations 
of Ichimura and collaborators --- without RPA correlations --- seem to 
be in good agreement with the experimental ratio at all three values of 
$q$. In particular, these trends suggest that relativistic RPA calculations 
with distortions could yield a good description of the momentum-transfer 
dependence of the ratio. It is clear, however, that before proceeding 
further with any theoretical comparison one must understand the interplay 
between ``mundane'' effects, such as distortions, and ``novel'' effects, 
such as meson-mass rescaling and/or relativity, in the determination of the 
isovector response.

There are, however, alternative ways of testing the different
theoretical models. For example, one could concentrate on the 
individual spin-longitudinal and spin-transverse responses, rather 
than on their ratio. This has the advantage that the transverse
response, even in plane-wave, can be compared directly to 
electron-scattering data. Since in the rescaling model the transverse
component of the interaction becomes attractive at 
$q\sim 2.2~{\rm fm}^{-1}$, while it remains repulsive in the relativistic 
model, one could compare readily the (different) predictions of the 
two models with existing electron-scattering data~\cite{barr83}.

Finally a word on future work.
One of the early indications of a diminished role of isovector 
correlations in the medium came from the measurement of $R_{L}/R_{T}$ 
in quasielastic $(\vec{p},\vec{p}')$ scattering~\cite{carey84,rees86}. 
Unfortunately, uncertainties associated with the removal of the isoscalar 
contribution clouded the interpretation. The $(\vec{p},\vec{n})$ reaction,
on the other hand, is free from any isoscalar contamination and was 
advertised as the most promising method of observing the predicted 
enhancement of $R_{L}/R_{T}$. Thus, the advent of new experimental 
facilities and techniques opened the door to precision studies
of the isovector response. Indirectly, these new facilities also 
opened the door to precision studies of the isoscalar response. Indeed,
combined $(\vec{p},\vec{p}')$ and $(\vec{p},\vec{n})$ data --- which
now exist --- should enable one, in principle, to extract the isoscalar 
spin-independent response $R_{0}$. This response is interesting 
because of its connection to the charge response measured in electron 
scattering and, thus, to the long-standing problem of the quenching of 
the Coulomb sum~\cite{barr83}. The charge response in electron 
scattering is dominated by the proton response which is half isoscalar 
and half isovector. Thus, electromagnetic probes are unable to isolate 
the pure isoscalar contribution to the response. Therein lies the appeal 
of the hadronic reactions.

Relativistic models of the electromagnetic response predict a 
substantial quenching of the charge response arising from 
isoscalar correlations~\cite{horpie89}. Indeed, relativistic 
effects reduce the isoscalar charge response by as much as 50\% 
relative to its Fermi-gas value. This large isoscalar quenching, 
however, is partially diluted by the isovector contribution to the 
electromagnetic response. Still, relativistic RPA results are 
about 25\% below the Fermi-gas response and in good agreement 
with experiment. In the case of hadronic probes the surface-peaked 
nature of the reaction forces the probe to sample a lower-density 
region than in electron scattering and should suppress some of the 
large ($\sim$50\%) isoscalar quenching. Still, a large reduction in 
$R_{0}$ appears to be an unavoidable consequence of the relativistic 
dynamics. A quantitative study of this effect is currently under 
investigation.

\acknowledgments

We would like to thank X.Y. Chen and T.N. Taddeucci for
making available the experimental data and for many helpful
discussions. This research was supported by the Florida State 
University Supercomputer Computations Research Institute through 
the U.S. Department of Energy contracts \# DE-FC05-85ER250000 
and \# DE-FG05-92ER40750.

%------------------- References --------------------

%
%--------------------- Figures ---------------------
\begin{figure}
 \caption{Feynman diagrams representing the RPA polarizations 
          and Dyson's equation for the isovector interaction.}
 \label{figone}
\end{figure}
\begin{figure}
 \caption{Effective $\pi$ and $\rho$ contributions
          to the residual $M^{\star}=M$ interaction 
          after the inclusion of the Landau-Migdal 
          parameter. The arrows are located at
          $q=1.2, 1.72,$ and $2.5$~fm$^{-1}$.} 
 \label{figtwo}
\end{figure}
\begin{figure}
 \caption{Transverse response for ${}^{40}$Ca$(e,e')$
          at $q=410$ and $550$~MeV. The dotted line
          is the Hartree response calculated in the
          finite nucleus. Nuclear-matter results with  
          (solid line) and without (dashed line)
          RPA correlations are also displayed.}
 \label{figthree}
\end{figure}
\begin{figure}
 \caption{Effective $\pi$ and $\rho$ contributions
          to the residual $M^{\star}\ne M$ interaction 
          after the inclusion of the Landau-Migdal 
          parameter. The arrows are located at 
          $q=1.2, 1.72,$ and $2.5$~fm$^{-1}$.}
 \label{figfour}
\end{figure}
\begin{figure}
 \caption{Spin-transfer combinations $(P-A_{y})$ 
          and $[(D_{S'L}+D_{L'S}) - (D_{L'L}+D_{S'S})
          \tan(\theta_{\rm lab})]$ for 
          ${}^{40}$Ca$(\vec{p},\vec{n})$ at $q=1.72$~fm$^{-1}$.
          Both combinations vanish in elastic scattering.}
 \label{figfive}
\end{figure}
\begin{figure}
 \caption{Spin-longitudinal and spin-transverse response
          functions calculated from nuclear matter with
          and without the inclusion of RPA correlations. 
          The response functions were, either, reconstructed
          from the spin-transfer observables or directly
          computed from appropriate polarization insertions.
          The quantities in square brackets give the value
          of the integrated response.}
 \label{figsix}
\end{figure}
\begin{figure}
 \caption{Cross section (divided by its single-nucleon value),
          and out-of-plane spin observables for 
          ${}^{40}$Ca$(\vec{p},\vec{n})$ at $q=1.72$~fm$^{-1}$. 
	  The dotted (dashed) line displays the 
          uncorrelated (RPA) Fermi-gas result.
          The dot-dashed (solid) line gives the uncorrelated
          (RPA) result using medium-modified nucleon masses.}
 \label{figseven}
\end{figure}
\begin{figure}
 \caption{In-plane spin-transfer observables for 
          ${}^{40}$Ca$(\vec{p},\vec{n})$ at $q=1.72$~fm$^{-1}$. 
	  The dotted (dashed) line displays the 
          uncorrelated (RPA) Fermi-gas result.
          The dot-dashed (solid) line gives the uncorrelated
          (RPA) result using medium-modified nucleon masses.}
 \label{figeight}
\end{figure}
\begin{figure}
 \caption{Polarization-transfer observables divided by their
          single-nucleon value for 
          ${}^{40}$Ca$(\vec{p},\vec{n})$ at $q=1.72$~fm$^{-1}$.
	  The dotted (dashed) line displays the 
          uncorrelated (RPA) Fermi-gas result.
          The dot-dashed (solid) line gives the uncorrelated
          (RPA) result using medium-modified nucleon masses.}
 \label{fignine}
\end{figure}
\begin{figure}
 \caption{Spin-longitudinal to spin-transverse ratio for
          ${}^{40}$Ca$(\vec{p},\vec{n})$ at $q=1.72$~fm$^{-1}$.
	  The dotted (dashed) line displays the 
          uncorrelated (RPA) Fermi-gas result.
          The dot-dashed (solid) line gives the uncorrelated
          (RPA) result using medium-modified nucleon masses.}
 \label{figten}
\end{figure}
\begin{figure}
 \caption{Spin-longitudinal to spin-transverse ratio for
          ${}^{12}$C$(\vec{p},\vec{n})$ at 
          $q=1.2, 1.72, 2.5$~fm$^{-1}$.
	  The dotted line displays the Fermi-gas result.
          The dot-dashed (solid) line gives the uncorrelated
          (RPA) result using medium-modified nucleon masses.}
 \label{figeleven}
\end{figure}
%
%--------------------- Tables ---------------------
 \mediumtext
 \begin{table}
  \caption{Meson masses, coupling constants, tensor-to-vector ratio,
           and Landau-Migdal parameter $g'$ for the isovector
           interaction. The value of $g'$ enclosed(not enclosed) in
           parenthesis should be used when 
           $M^{\star}\ne M(M^{\star}= M)$.}
   \begin{tabular}{ccccc}
    Meson & Mass(MeV) & ${g^2/4\pi}$ & $C=f/g$ & $g'$   \\
        \tableline
    $\pi $    &  139  &  14.08 &  ---  &  0.70 (0.70)   \\
    $\rho$    &  770  &   0.41 &  6.1  &  0.65 (0.30)   \\
   \end{tabular}
  \label{tableone}
 \end{table}

 \mediumtext
 \begin{table}
  \caption{Average Fermi momenta and effective masses for 
           an incident energy of $T_{\rm lab}=495$ MeV.}
   \begin{tabular}{cccc}
    Target & $k_{F}({\rm fm}^{-1})$ & $M^{\ast}_{p}/M$ & $M^{\ast}_{t}/M$ \\
        \tableline
    ${}^{12}$C  &  0.91  &  0.91  &  0.87  \\
    ${}^{40}$Ca &  0.94  &  0.90  &  0.85  \\
   \end{tabular}
  \label{tabletwo}
 \end{table}

 \mediumtext
 \begin{table}
  \caption{Spin-transfer observables from the
           ${}^{2}$H$(\vec{p},\vec{n})$ reaction
           at $q=1.72$~fm$^{-1}$ compared to the
           free NN values (obtained from the 
           nuclear-matter calculation in the 
           $k_{F} \rightarrow 0$ limit).}
   \begin{tabular}{ccccccc}
     $A_{y}$ & $P$ & $D_{NN}$ &  $D_{S'S}$ & $D_{S'L}$ &
                     $D_{L'S}$ & $D_{L'L}$                      \\
    $ 0.13 \pm 0.00$ & $ 0.12 \pm 0.01$ & $ 0.01 \pm 0.03$ &
    $-0.21 \pm 0.03$ & $-0.11 \pm 0.03$ & $-0.03 \pm 0.03$ &
    $-0.47 \pm 0.03$                                            \\
    $ 0.14 $ & $ 0.14 $ & 
    $-0.15 $ & $-0.20 $ & 
    $-0.14 $ & $ 0.03 $ & 
    $-0.52 $ \\
   \end{tabular}
  \label{tablethree}
 \end{table}

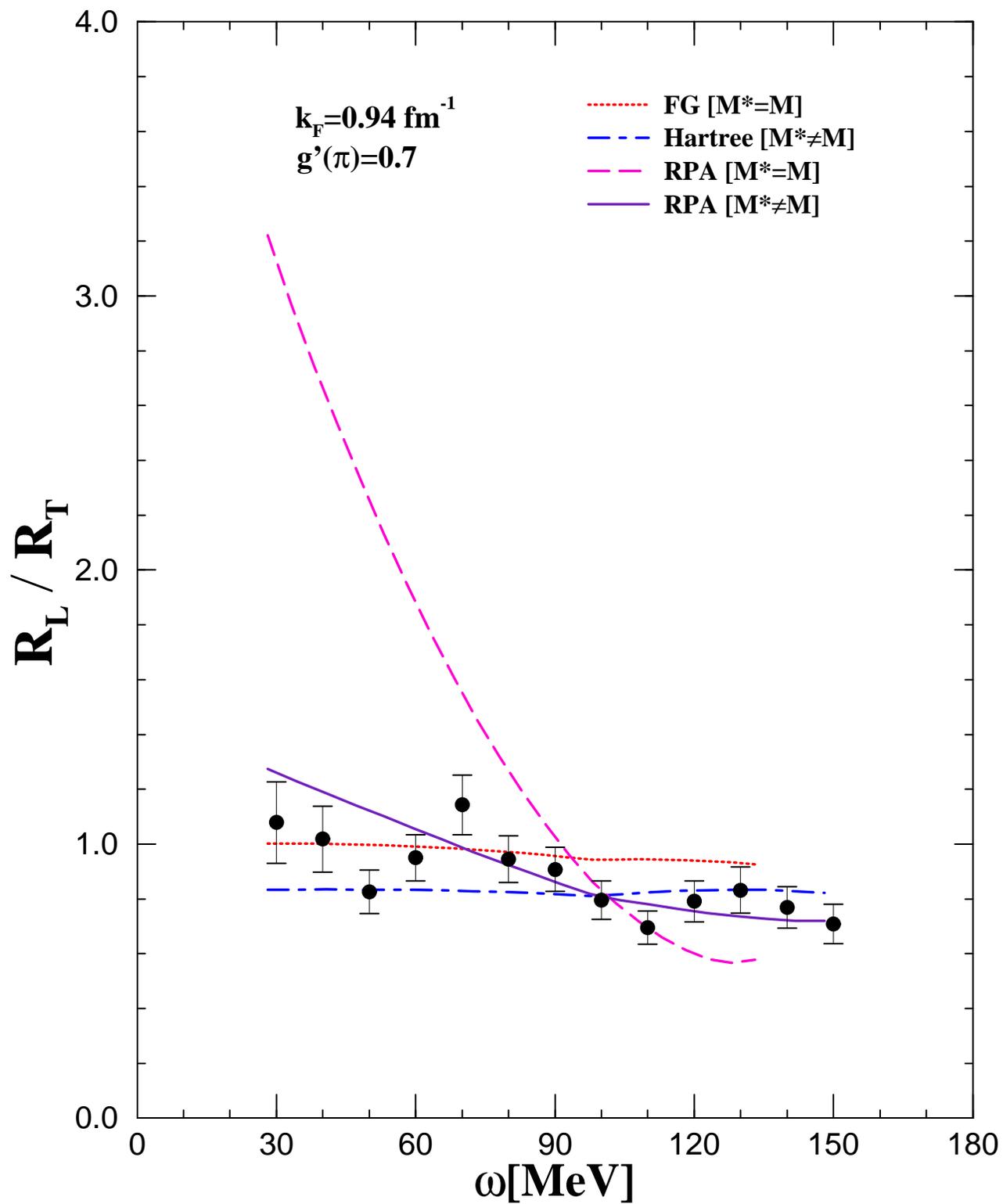
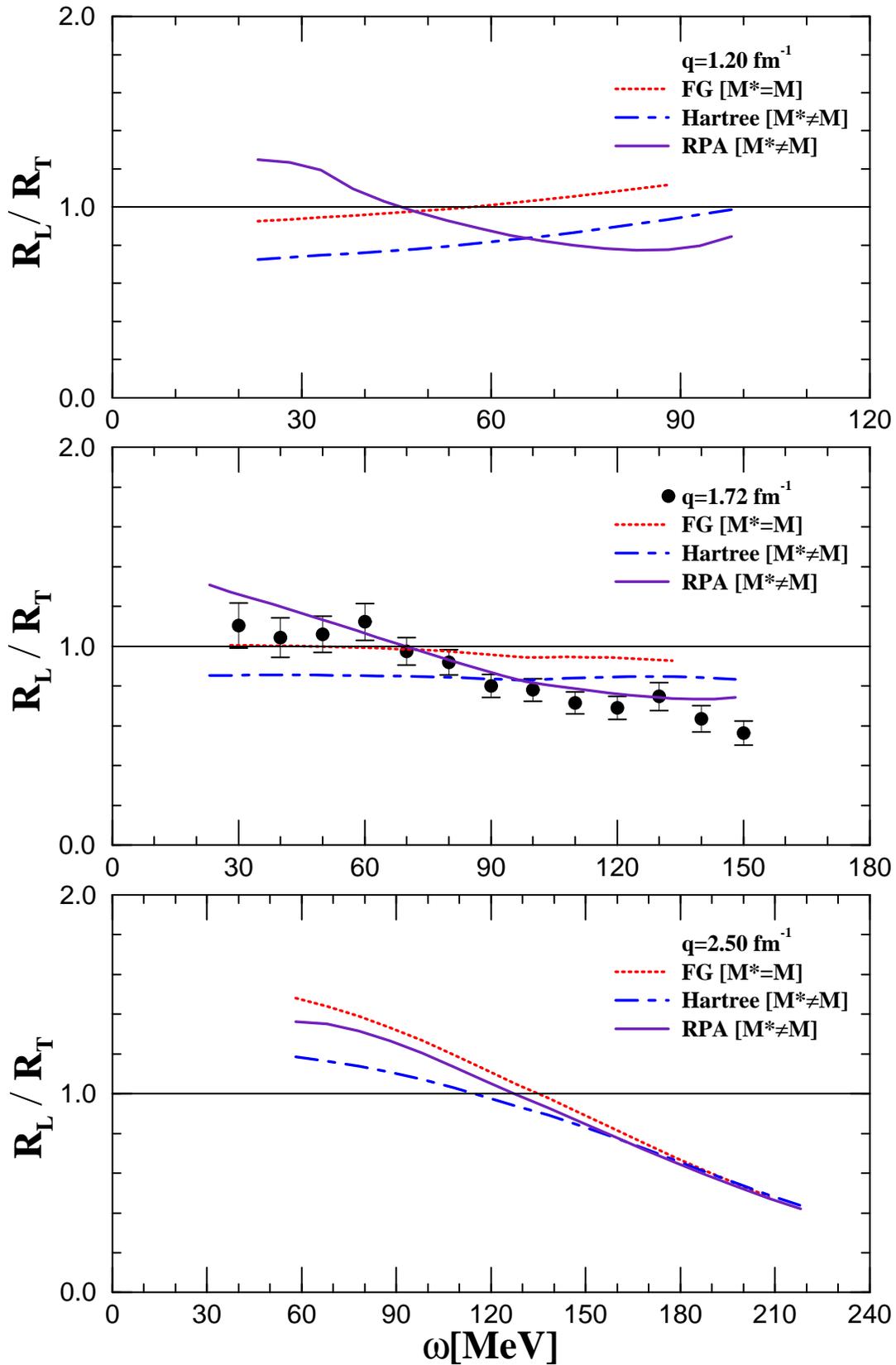
\end{document}